\documentclass[twoside]{dis09}
\usepackage[latin1]{inputenc}
\usepackage[dvips]{graphicx,epsfig,color}
\usepackage{wrapfig,rotating}
\usepackage{amssymb,amsmath,array}

\begin{document}
\title{Searches for New Neutral Gauge Bosons and Leptoquarks at the Tevatron}

\author{Alexei N. Safonov \\
(For CDF and D0 Collaborations)
%
\vspace{.3cm}\\
%
Texas A\&M University - Department of Physics \\
College Station, TX 77843 - USA
}

\maketitle

\begin{abstract}
This contribution reports on some of the most recent searches for new heavy
neutral bosons and leptoquarks performed at the Tevatron experiments.
\end{abstract}

\section{Introduction}

Despite of its tremendeous success in describing wealth of existing experimental data, 
the Standard Model (SM) has significant shortcomings, e.g. the hierarchy problem and the
failure to explain origin of the electroweak symmetry breaking, matter-antimatter 
asymmetry, and apparent presence of dark matter. Many of the extensions proposed to
resolve some or most of these problems predict the existence of new heavy particles.
Heavy neutral bosons appear in many models, e.g. $Z^\prime$'s 
appearing in the string-inspired E6 models~\cite{wprime_zprime}, Randall-Sundrum (RS) 
gravitons~\cite{RS_graviton}, heavy new bosons appearing in ``little higgs''~\cite{littleHiggs} 
and left-right symmetric models~\cite{leftrightsymmodels}, sneutrinos in R-parity 
violating SUSY\cite{snu_susy}, as well as strongly interacting excited axigluons~\cite{axigluon}, 
colorons~\cite{coloron}, techni-$\rho$'s~\cite{technirho}. 
Similarly, leptoquarks appear in models intending to explain the apparent lepton-quark symmetry of 
the SM~\cite{lqs}. Another example is SUSY with R-parity violation where a leptoquark role can be taken
by scalar quarks. While different models have varying predictions for the new particle 
production and decay mechanisms and dynamics, they all share similar experimental 
signatures and have been the subject of exhaustive experimental searches at both 
CDF and D0 experiments at the Tevatron. This contribution reviews some of the most
recent of those searches.

\section{New Boson Searches in Dilepton and Dijet Channels}

\begin{figure}[t]
\centerline{\includegraphics[width=0.42\columnwidth]{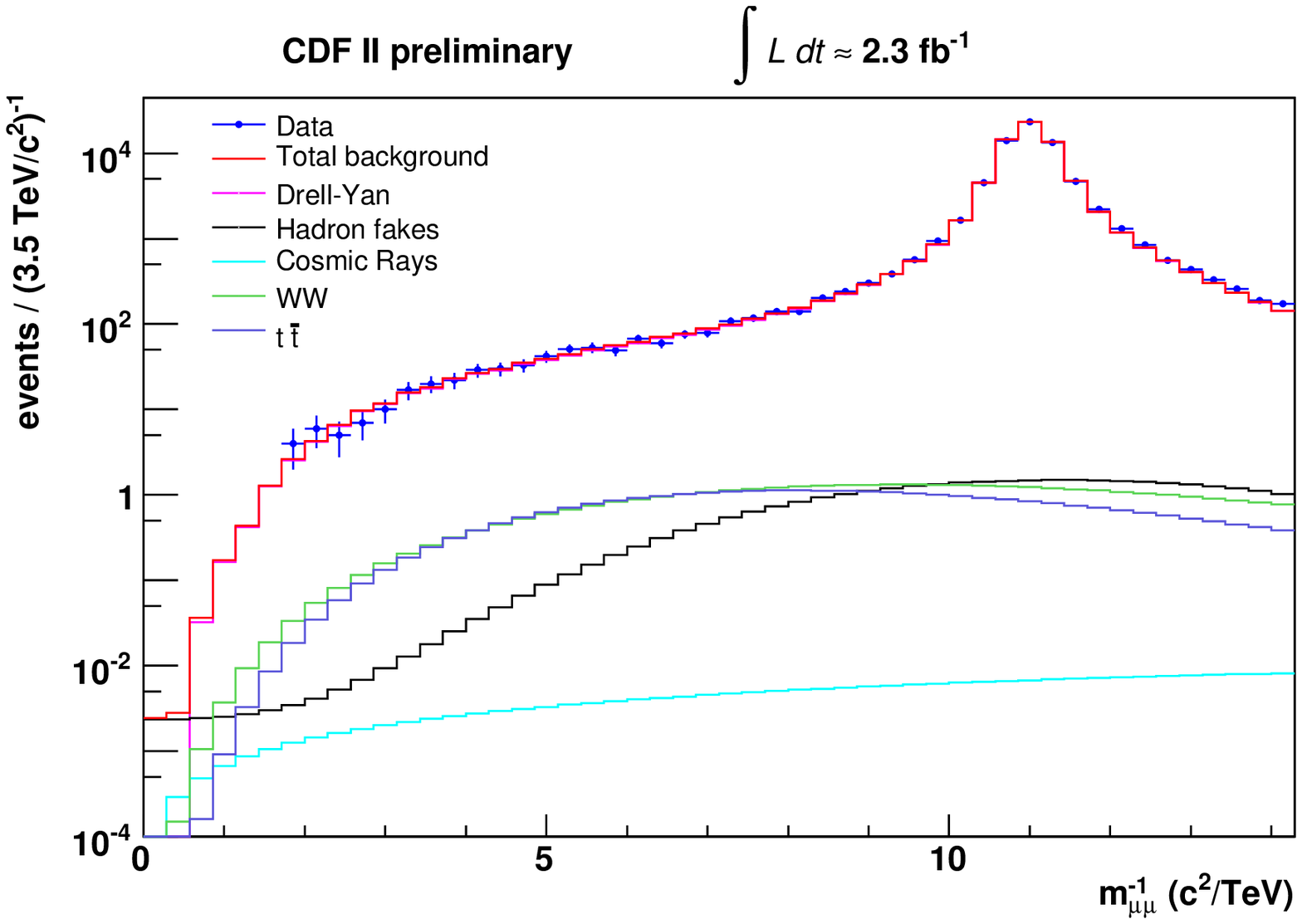}
\includegraphics[width=0.48\columnwidth]{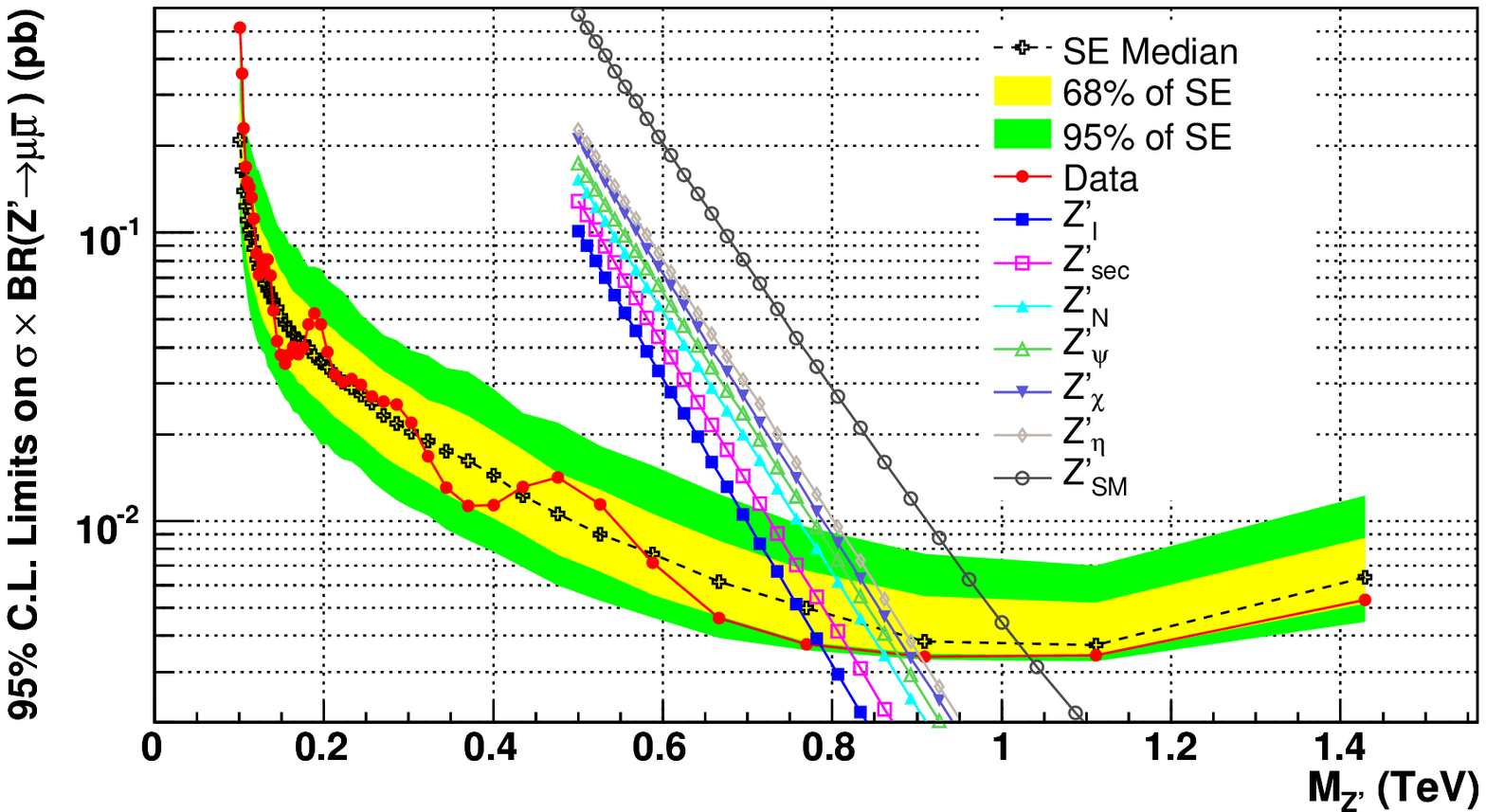}}
\caption{Invariant mass distribution of muon pairs and corresponding limit on Z-prime}\label{Fig:dimuon_figure}
\end{figure}

In many schemes of GUT symmetry-breaking, U(1) gauge groups survive to relatively low energies~\cite{lowmassU1_GUT}, leading to the 
prediction of neutral gauge vector bosons, generically referred to as Z$^\prime$ bosons. Such Z$^\prime$ bosons typically 
couple with electroweak strength to SM fermions, and can be observed at hadron colliders as narrow, spin-1, dilepton resonances
from $q\bar{q} \to Z^\prime \to l^+l^-$. Many other SM extensions, such as the  left-right symmetric~[5] and the ``little Higgs'' 
models~\cite{littleHiggs}, also predict heavy neutral gauge bosons. Additional spatial dimensions are a possible explanation for the gap between the 
electroweak symmetry breaking scale and the gravitational energy scale $M_{Planck}$. In the Randall-Sundrum (RS) scenario~\cite{RS_graviton},
the ground-state wave function of the graviton is localized on a three-dimensional ``brane'' separated in a fourth
spatial dimension from the SM brane. The wave function varies exponentially in this fourth dimension, causing its
overlap with the SM brane to be suppressed and explaining the apparent weakness of gravity and the large value
of $M_{Planck}$. This model predicts excited Kaluza-Klein modes of the graviton which are localized on the SM
brane. These modes appear as spin-2 resonances $G^*$ in the process $ q\bar{q} \to G^* \to l^+l^-$, with a narrow intrinsic width
when $k/M_{Planck} < 0.1$, where $k^2$ is the spacetime curvature in the extra dimension. Spin-0 resonances 
such as the sneutrino $\tilde{\nu}$ in the process
$q\bar{q} \to \tilde{\nu} \to l^+l^-$ are predicted by supersymmetric theories with R-parity violation~\cite{snu_susy}. 
Experimentally, these new bosons are sought by looking for narrow resonances in di-electron or di-muon spectra.
These final states are nearly free of instrumental backgrounds and are dominated by the 
irreducible Drell-Yan contribution. CDF has recently published results on searches 
for $Z^\prime$ and ED graviton in di-electron and di-muon channels~\cite{CDF_ee_mm} using 2.5 $fb^{-1}$ 
and 2.3 $fb^{-1}$ of data, respectively. Analysis in the muon channel requires two 
oppositely charged tracks with $p_T>30$ GeV/c, consistent with the hypothesis that they are minimum ionizing
particles. Figure~\ref{Fig:dimuon_figure} 
shows the invariant mass distribution demonstrating impressive agreement between data and the SM expectation. 
With no excess, the cross-section limit is calculated and is shown in Fig.~\ref{Fig:dimuon_figure}
for different $Z^\prime$ species. For RS $G^*$, the limit is $G^*>921$ GeV/c$^2$ for $k/M_{Planck} = 0.1$. 
A similar search in the di-electron channel yields $m(G^*)>807$ GeV/c$^2$. 

\begin{figure}[t]
\centerline{\includegraphics[width=0.53\columnwidth]{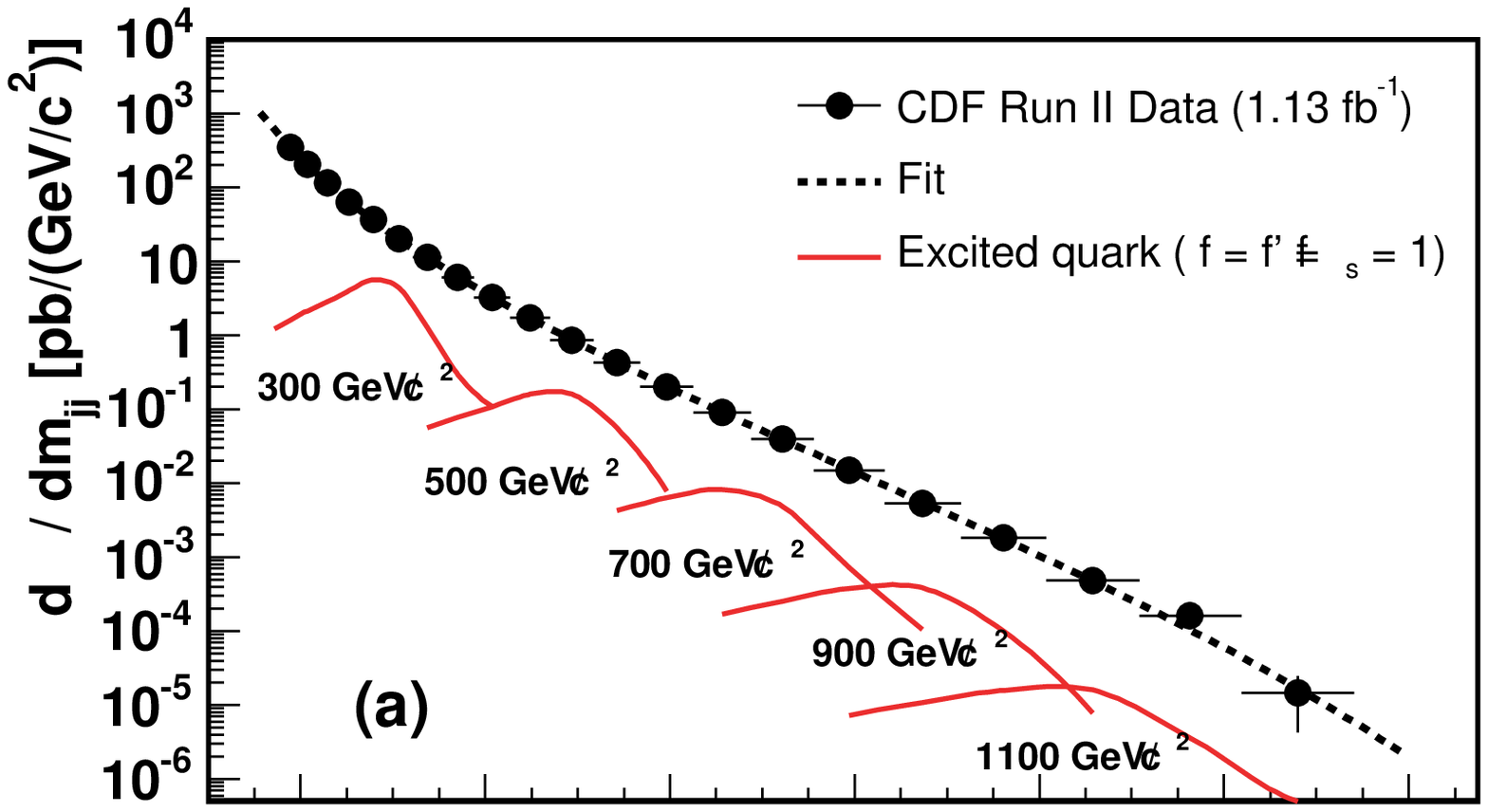}
\includegraphics[width=0.37\columnwidth]{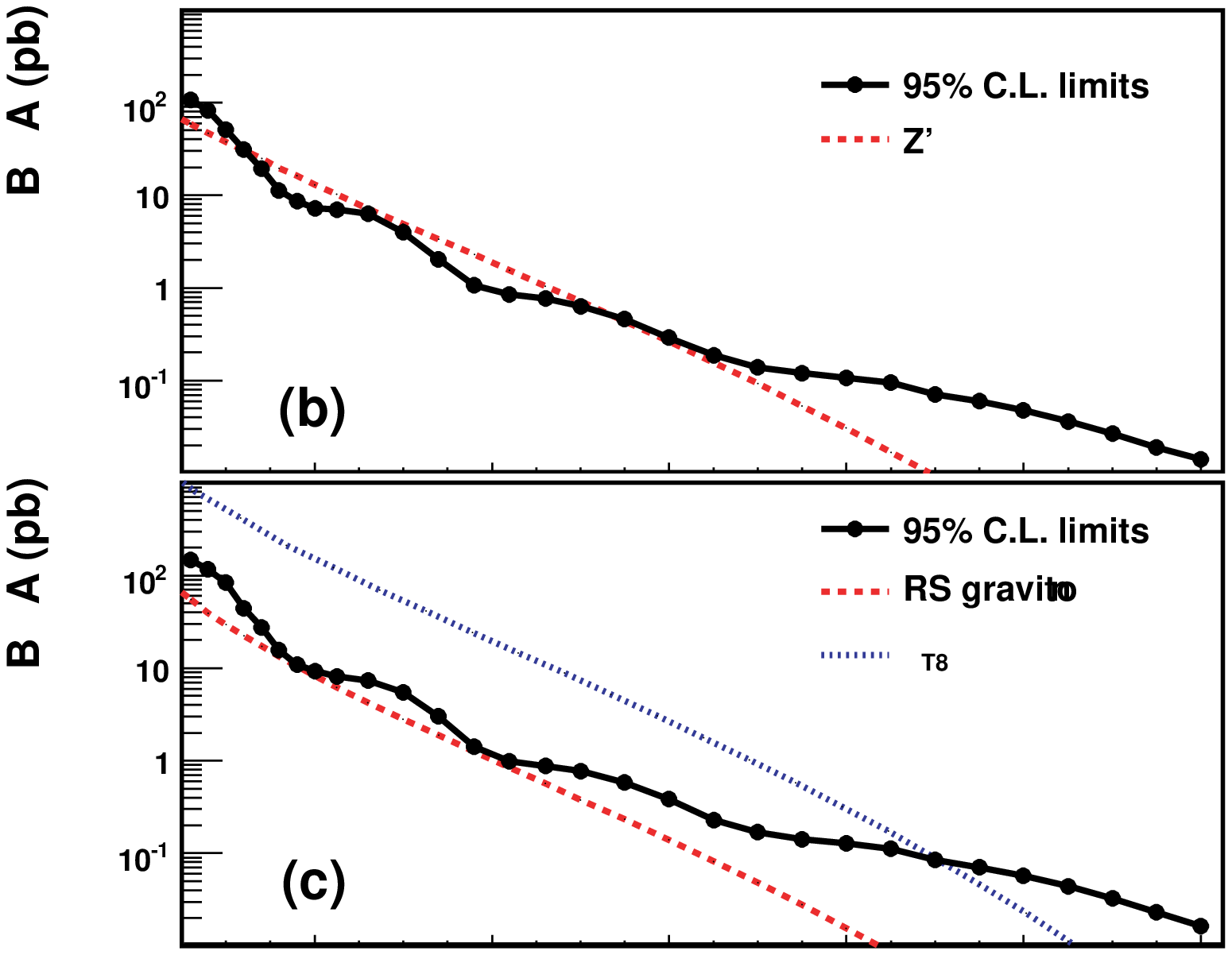}}
\caption{Dijet invariant mass spectrum and the corresponding cross-sections versus new particle mass plots
showing exclusion levels for several scenarios.}\label{Fig:dijet_figure}
\end{figure}

There are several compelling scenarios motivating searches
for heavy new resonances decaying to quarks and gluons. In chiral color models, the SU(3) gauge group of QCD 
results from the spontaneous breaking of the chiral color gauge group of $SU(3) \times SU(3)$, leading to the 
presence of the axigluon, a massive axial vector gluon, that decays to $q\bar{q}$~\cite{axigluon}. 
The $E_6$ GUT model also predicts the presence of a diquark which decays to $qq$ or $\bar{q}\bar{q}$~\cite{GUT_diquark}.
The flavor-universal coloron model~\cite{coloron} predicts the presence of a color-octet coloron which decays to 
$q\bar{q}$, models of extended technicolor and topcolor-assisted technicolor~\cite{technirho} predict the 
presence of a color-octet techni-$\rho$ ($\rho_{T8}$) decaying to $q\bar{q}$ or $gg$. Together with dilepton 
analyses, searches with dijets improve sensitivity to the models predicting $Z^\prime$ bosons and
ED gravitons discussed earlier. In the case of RS model, dijet channels may have special importance as the effective coupling of $G^*$ to 
the SM particles can be enhanced or suppressed depending on their localization~\cite{modifiedRS}. A recent CDF study~\cite{CDF_jj} examines the 
dijet spectrum looking for evidence of a new resonance on top of the QCD dijet spectrum using 1.1 $fb^{-1}$ of data 
collected using jet triggers. After removing instrumental backgrounds, e.g. beam losses and cosmic rays, jets are 
corrected for non-uniformities and non-linearities in the calorimeter response, and the two highest $E_T$ jets are 
used to calculate the dijet mass $m_{jj}$. After correcting for smearing due to calorimeter resolution and offline 
selection requirements, the resulting $m_{jj}$ spectrum is shown in Fig.~\ref{Fig:dijet_figure}. The search for 
dijet mass resonances is performed by parameterizing the shape of the dijet distribution with a smooth function 
and fitting data for statistically significant deviations consistent with the new dijet resonances. Figure~\ref{Fig:dijet_figure} 
depicts the expected shape of the ``bump'' due to a new physics resonance using an excited 
quark model as an example. Note that the shape is nearly independent of the type of resonance as it is dominated by the 
calorimeter resolution. With no excess, the  data are used to restrict allowed cross-section for new 
particle production in several new physics scenarios as well as the new particle masses. Figure~\ref{Fig:dijet_figure} shows corresponding exclusion
plots for benchmark $Z^\prime$ and RS graviton $G^*$ scenarios. The search excludes $260 < m < 1250$ GeV/c$^2$  for the axigluon and flavor-universal coloron, 
$290 < m < 630$ GeV/c$^2$ for the E6 diquark, $260 < m < 1100$ GeV/c$^2$ for $\rho_{T8}$.

\section{New Boson Searches in Di-Boson Final States}

\begin{figure}[t]
\centerline{\includegraphics[width=0.43\columnwidth]{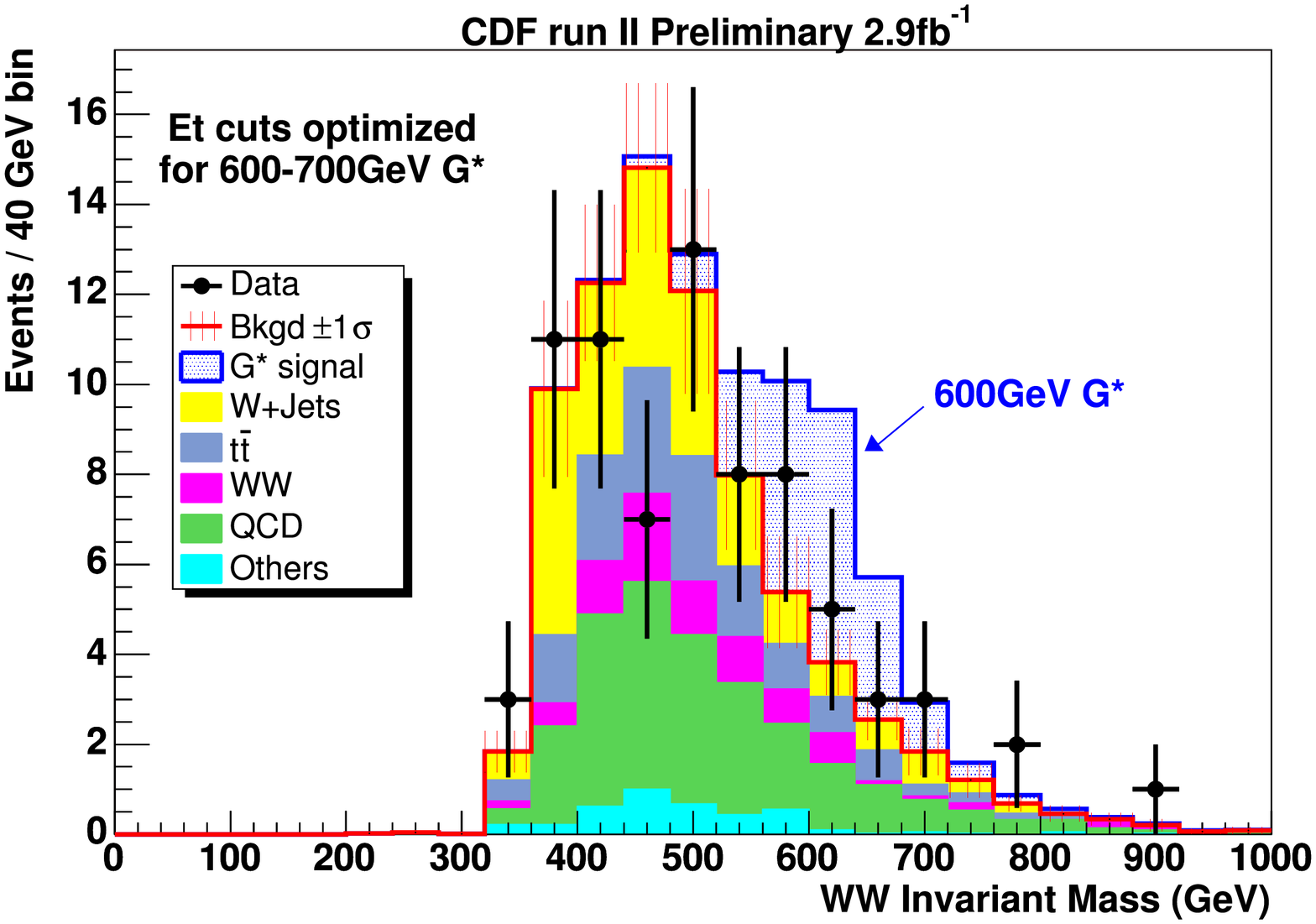}
\includegraphics[width=0.44\columnwidth]{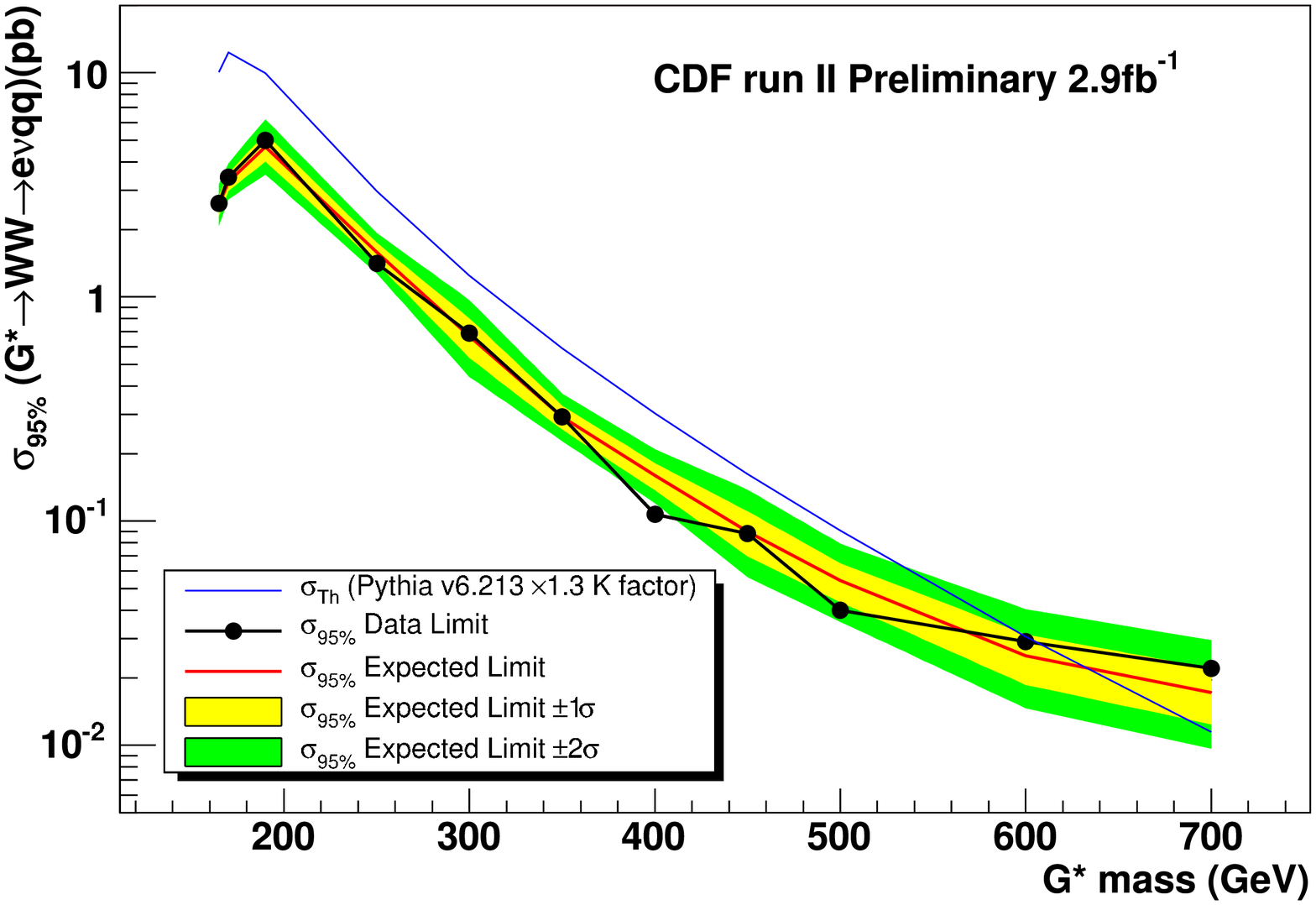}}
\caption{Invariant mass of $WW$ candidate events and the 95\% C.L. limit on $\sigma (p\bar{p} \to G^*)$ as a function of $m(G^*)$.}\label{Fig:ww_figure}
\end{figure}

Searches for new heavy bosons decaying to a pair of SM gauge bosons, e.g. $WW$ or $ZZ$, can provide an important complementarity
to the dilepton channels. Furthermore, if for some reason the coupling of new bosons to fermions is suppressed, the di-boson mode
can hold the key to discovery of new physics, e.g. as modified RS scenarios discussed earlier where SM particles can be in the bulk, e.g. leading to reduced 
effective couplings of $G^*$ to leptons~\cite{modifiedRS}. CDF has recently performed a search~\cite{CDF_jj} for heavy new resonance decaying to $WW$ using
2.9 $fb^{-1}$ of data. The search is performed using the $ejj$+MET final state, where one of the $W$'s decays leptonically
and the other one is allowed to decay into a pair of jets to enhance the acceptance of the analysis compared to the purely 
leptonic mode. Selected events are required to have an isolated electron ($E_T > 30$ GeV), a missing $E_T > 30$ GeV, 2 or 3 
jets with $| \eta |< 2.5$ and $ E_T > 30$ GeV, and an overall $H_T > 150$ GeV. $H_T$ is defined as the sum of the electron $E_T$, 
the missing $E_T$ and the jet $E_T$ of all jets in the event. To reconstruct the $WW$ topology, the electron and missing 
$E_T$ are used to solve for missing $E_Z$ under the assumption that the electron momentum and missing energy comprise the $W$ 
mass. The event is dropped if no solution is possible. Events are further required to have a pair of jets with the 
di-jet invariant mass consistent with the $W$ mass within resolution. The final step is to optimize selections for 
either the $G^*$ or SSM $Z^\prime$ hypothesis, and separately for several possible mass ranges of the new boson. This is achieved
by selecting sub-samples with varying minimum thresholds for missing $E_T$, lepton and jet $E_T$. The reconstructed invariant 
mass of candidate events for one of such sub-samples is shown in Fig.~\ref{Fig:ww_figure}. With no statistically significant
excess of data over SM, limits are set on the production cross-section of $Z^\prime$ and $G^*$. Using standard RS gravition 
and SSM $Z^\prime$ as a benchmark, the excluded mass ranges are $m(G^*)<607$ and $247<m(Z^\prime)<545$ GeV/c$^2$.

\section{Searches For Leptoquarks}

\begin{figure}[t]
\centerline{\includegraphics[width=0.40\columnwidth]{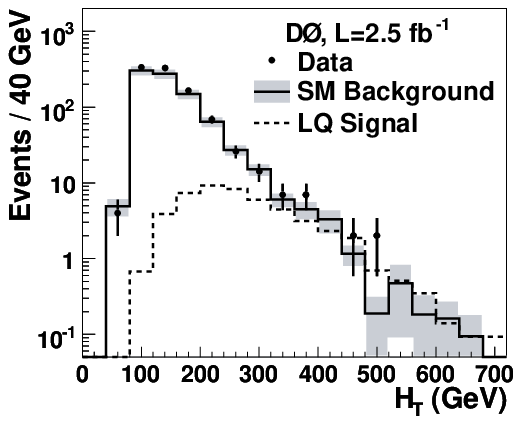}
\includegraphics[width=0.47\columnwidth]{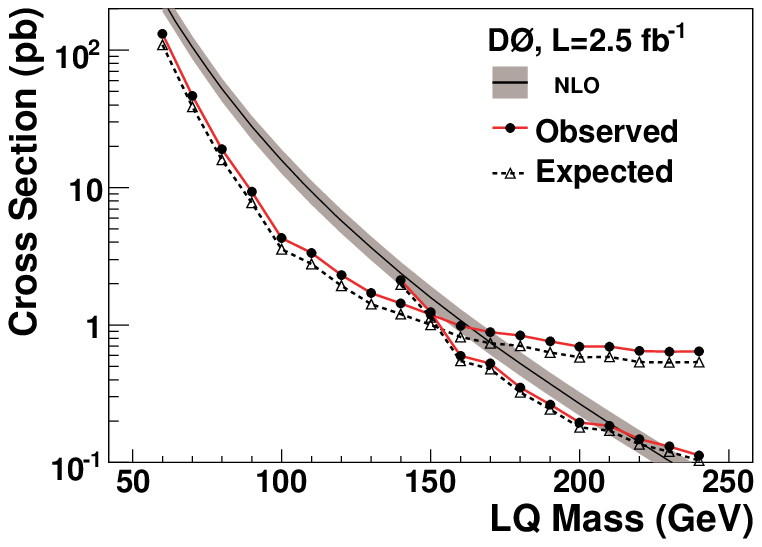}}
\caption{Distribution of $H_T$ for selected candidate events and the corresponding 95\% C.L. limit on 
$\sigma(p \bar{p} \to LQ)$ vs leptoquark mass.}\label{Fig:lqjetmet_figure}
\end{figure}

Models attempting to explain the symmetry of the lepton and quark sectors in the SM often predict the existence of 
leptoquarks (LQ)~\cite{lqs}. Those are scalar or vector particles carrying both a lepton and a baryon 
quantum number. At hadron colliders, new colored particles predicted by various extensions of the standard model 
(SM) would be abundantly produced if they are light enough. To satisfy experimental constraints on changing neutral 
current interactions, leptoquarks couple only within a single generation. Leptoquarks decay into a charged lepton 
and a quark with a branching ratio $\beta$, or into a neutrino and a quark with a branching ratio $\beta -1$. 
Pair production of leptoquarks assuming $\beta = 0$ therefore leads only to a final state consisting of two neutrinos 
and two quarks. The corresponding experimental signature is the presence of jets and missing transverse energy 
resulting from the decay of those particles. A recent D0 analysis~\cite{D0_lqs} explores the $jj+$missing $E_T$  channel 
by analyzing events with the topology consisting of exactly two jets and missing $E_T$ using 2.5 $fb^{-1}$ of data. 
Because the final state has no leptons, the analysis has similar acceptance to leptoquarks belonging to any of the 
three possible generations. Before final optimization, the event selection requires presence of exactly two acoplanar 
jets ($\Delta \phi (j_1,j_2)<170 ^o$) with $E_T>35$ GeV in the central part of the detector and missing $E_T$ over 75 
GeV. To minimize instrumental mismeasurements of missing energy and suppress large QCD multi-jet background 
contamination, the missing $E_T$ in selected events is required to point away from any 
of the jets in the event. Events containing identified lepton or isolated track candidates are removed to reduce the
$W+$jet and $t\bar{t}$ backgrounds. Figure~\ref{Fig:lqjetmet_figure} shows the distribution of $H_T$ defined as a scalar sum 
of the transverse energies of the jets in the event and the missing $E_T$. At this stage, the analysis is
split into two separate searches targeting signals of either lower or higher leptoquark mass. Best sensitivity
to lighter leptoquark signal is achieved by additionally requiring $H_T>150$ GeV, while the higher mass search requires
$H_T>300$ GeV and uses a tighter cut on missing $E_T$ of 125 GeV. Neither of the two sub-analyses found significant
excesses of data over the SM expectation. The 95\% C.L. upper bound on the leptoquark production 
cross-section for $\beta =0$ is shown in Fig.~\ref{Fig:lqjetmet_figure}. The corresponding leptoquark mass limit is 
205 GeV/c$^2$ using NLO predicted cross-section~\cite{lqs}. 

\section{Bibliography}
\begin{footnotesize}

\end{footnotesize}


\end{document}